# A Study on Application of Spatial Data Mining Techniques for Rural Progress

- V. R. Kanagavalli[1], Research Scholar, Sathyabama University, Chennai
- Prof. Dr. K. Raja[2], Principal, Narasu's Sarathy Institute of Technology, Salem.

**Abstract**

**This paper focuses on the application of Spatial Data mining Techniques to efficiently manage the challenges faced by peripheral rural areas in analyzing and predicting market scenario and better manage their economy. Spatial data mining is the task of unfolding the implicit knowledge hidden in the spatial databases. The spatial Databases contain both spatial and non-spatial attributes of the areas under study. Finding implicit regularities, rules or patterns hidden in spatial databases is an important task, e.g. for geo-marketing, traffic control or environmental studies. In this paper the focus is on the effective use of Spatial Data Mining Techniques in the field of Economic Geography constrained to the rural areas.**

## 1. Introduction

### 1.1 Economic geography

Economic Geography is the study of the location, distribution and spatial organization of economic activities across the Earth.

One of great contribution of the New Economic Geography (NEG) is to explicitly model "the self-reinforcing character of spatial concentration" (Fujita, Krugman and Venables 1999 p.4).

The concept of Economic Geography directly relates to the Tobler's first law of geography which states that nearby entities often share more similarities than entities which are far apart.

Study of Agglomeration Economics, Transportation, Real Estate, Gentrification, Ethnic Economies, Gendered Economies, Globalization, Location of Industries etc depends on the knowledge of Economic Geography.

In a nutshell, the main objective of Economic Geography is to
- Focus on industrial location and use quantitative methods
- take into account social, cultural, and institutional factors in the spatial economy

Unlike an Economist, an economic geographer will use his expertise in many fields to determine the underlying causes of an economic problem holistically.

### 1.2 Spatial Data Mining Techniques

**Spatial patterns** are most often the effect of some kind of influence of an object on other objects in its neighborhood which typically decreases or increases more or less continuously with increasing or decreasing distance.

**Spatial Clustering** is the task of creation of thematic maps by clustering feature vectors. The constraint is to group the spatial objects into meaningful subclasses so that the members of a cluster are as similar as possible whereas the members of different clusters differ as much as possible from each other.

**Spatial Characterization** is the task of finding a compact description for a selected subset of the database. Again, finding a *spatial* characterization involves not only the properties of the target objects, but also the properties of their neighbors.

**Spatial Trend Detection** is defined as the process of finding a regular change of one or more non-spatial attributes when moving away from a given start object *o*. *Spatial trend*s describe a regular change of non-spatial attributes when moving away from a start object *o*.
A Spatial Trend may be Global or Local.

**Global Trends**, i.e. trends for the whole set of all neighborhood paths with source *o* having a length in the specified interval.

**Local Trends**, i.e. trends for a single neighborhood path with source *o* having a length in the specified interval

**Spatial classification** is the process of assigning an object to a class from a given set of classes based on the attribute values of the object. In *spatial classification* the attribute values of neighboring objects may also be relevant for the membership of objects and therefore have to be considered as well.

## 2. Literature Study

Many researches have studied the field of Economic Geography and the application of Spatial Data Mining Techniques in the development of Rural Economics.
The authors have studied the development of urban areas in many of the previous works. In [31], the author presents a Study on the effect of clustering of industries and a change in Economic Geography. [18] throws an insight into the growth of a city under study where pattern detection is explored. In [18], [19], [20], the authors study the application of the high resolution pictures available from SAR (Synthetic Aperture Radar) for the study of urban development. Eklund, Kirkby and Salim (1998) use inductive learning techniques and artificial neural networks to classify and map soil types. Lees and Ritman (1991) use decision tree induction methods for mapping vegetation types in areas where terrain and unusual disturbances (e.g., fire) confound traditional remote sensing classification methods.

## 3. Role of Economy in Development

### 3.1 Economic Growth

Economic growth has been defined by Arthur Lewis as *"the growth of output per head of population"*.

According to Arthur Lewis, economic growth is conditioned by

- Economic activity
- Increasing knowledge
- Increasing capital.

Putting it in more practical terms, they stand for the three factors, namely labour, technical improvements and capital.

An Economy would be balanced if there is a equal growth of all of all sectors, namely,
- Agriculture
- Manufacturing Industry and
- The Service Sector.

This uniform economic growth will benefit all sectors of the population.

### 3.2 Economic Development

Economic welfare of a nation depends not only on the growth of output but on the way it is distributed among different factors of production in the form of rent, wages, interest and profits.

A country is said to be Economically developed if there is
- Decline In Poverty,
- Decline in Unemployment and
- Decline in Inequality

But unfortunately even if per capita income doubled, there is no economic development in our country since there is very little improvement in the quality of life.

Most of the people in the rural areas are still to have higher incomes, better education, better health care and nutrition, less poverty and more equality of opportunity.

According Michael P. Todaro and Stephen C. Smith, "development must be conceived of as a multidimensional process involving major changes in social structures, popular attitudes and national institutions, as well as the acceleration of economic growth, the reduction of inequality, and the eradication of poverty".

## 5. The Current Rural Scenario

In [30], the experts analyse the various industries and resources of Indian Rural Economy and present views on improvement.

More efficient and competitive agricultural markets and agro-industries can deliver better prices and greater market opportunities to farmers without raising prices to consumers.

They can also expand employment opportunities in the non-farm economy by raising farm incomes and demand for rural goods and services, and by enabling the rural industry to modernize, compete better, and expand its domestic and export markets.

A look at the data nearly a decade before still holds true in this decade.

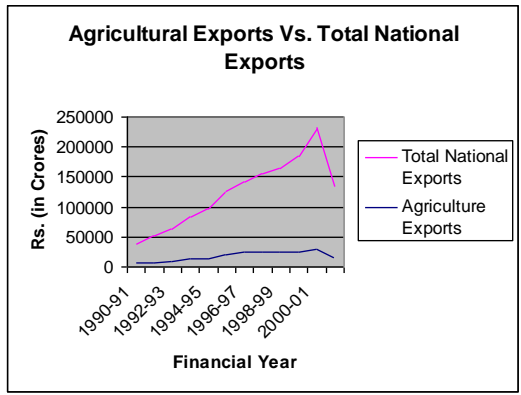

Fig. 1 Agricultural Exports Vs. Total National Exports

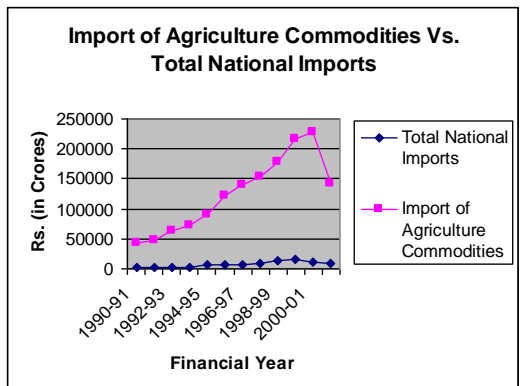

Fig. 1 Agricultural Imports Vs. Total National Imports

An integrated approach of improving the national agricultural exports and decreasing the agricultural imports would improve not only rural economy but would improve the overall economy of the nation.

## 5.1 Need for the Study

Rural Economy would go through uplift if necessary attention is put on the identification of different approaches to achieve improved quality of delivery. The research should focus on the improvement of the range of infrastructure and support services and also how they may vary across regions.

This research would definitely need the assistance of spatial data mining which would throw light on the spatial attributes of the various rural regions and also the non-spatial attributes such as the yield, soil nature, population size etc.,

The quality of the rural services would be promoted by devising an approach to promote a more participatory and demand-driven approach to the delivery of public goods and services.

## 6. Application of Spatial Data Mining Techniques in Rural Development

The Spatial Data Mining Techniques can be applied to achieve the desired Rural Economic Growth and Development.

### 6.1 Spatial Pattern Detection

The Spatial Pattern Detection can be used to find out how an economic recession in one sector affects the nearby rural area's performance in various sectors like education, agriculture, horticulture, health management systems etc.,

The patterns so predicted would help the analysts to concentrate more on the fields and to rectify or to necessary steps like providing more aids to reduce the risk of farmers facing more poorer living conditions.

### 6.2 Spatial Clustering

Spatial Clustering may be used to identify the density of rural industries in specified locations. This would help to find the availability of natural resources and technical skills of the people in a specified area.

Clustering techniques thus used may be used to put in more aid in that area and improve the marketing strategies to ensure better sales strategies.

Also, the clustering strategy may help to find the spread of epidemics, concentration of specific diseases in that particular area and thus perform a better root cause analysis for the repeated spread of disease in that particular area and take the needed remedial action.

### 6.3 Spatial Characterization

Spatial Characterization would help in finding the comparison between the performance of rural industries Vs. Urban industries, Literacy rates of Rural areas Vs. Neighboring non-rural areas, Spread of specific diseases in comparison to the urban locations in the past few years etc.,

### 6.4 Spatial Trend Detection

The Spatial Trend Detection is useful in finding the increase in literacy rate, increase in the agricultural yield, increase of use of pesticides; decrease/increase of male: female ratio etc., the information gained out of this trend detection would help for planning better campaigning programmes to enlighten the rural people in the respective areas.

### 6.5 Spatial Classification

Spatial Classification may help to find out the various kinds of occupation an area specializes. Even in agriculture it may help to find the differences in the kind of product an area specializes. This would help to plan for relevant industry setting up in a particular area.

Spatial Classification would help in deciding and implementing better transportation of one product to an area where it is to be deployed.

Spatial Data Mining Techniques also help to decide the distribution centers of the various resources.

### 7. Conclusion

Development of a state begins from the development of the rural districts of the state. Often the infrastructure facilities available in the town and other rural areas are not given much importance.

Rural development is concerned with economic growth and social justice, improvement in the living standards of the rural people by providing adequate and quality society services and minimum basic needs.

Rural infrastructure contributes more to the economic development of the country. It is important to bring those infrastructure details in limelight.

Spatial Data Mining Techniques provide a better understanding of the resources and shortcomings of rural area and help the administrative people to plan better strategies to deploy and distribute the resources better.

This paper aims at making the rural infrastructure data's beneficial for the society and has very good social relevance.